\documentclass[aoas,MSNbibl,nameyear,rotating,dvips]{arximspdf}
\usepackage{dcolumn,multirow}
\usepackage{graphicx}

\doi{10.1214/11-AOAS527} 
\volume{6}
\issue{2}
\pubyear{2012}
\firstpage{457}
\lastpage{475}

\makeatletter
\newcolumntype{d}[1]{D{.}{.}{#1}}
\makeatother

\begin{document}
\begin{frontmatter}

\title{Improving sequence-based genotype calls with linkage
disequilibrium and pedigree information\thanksref{T1}}
\runtitle{Improving sequence-based genotype calls}
\thankstext{T1}{Supported by NIH Grant CA094069.}

\begin{aug}
\author[A]{\fnms{Baiyu} \snm{Zhou}}
\and
\author[A]{\fnms{Alice S.} \snm{Whittemore}\corref{}\ead[label=e1]{alicesw@stanford.edu}}
\runauthor{B. Zhou and A. S. Whittemore}
\affiliation{Stanford University}
\address[A]{Department of Health Research and Policy\\
Stanford University\\
School of Medicine\\
Redwood Bldg, Room T204\\
Stanford, California 94305\\
USA\\
\printead{e1}} 
\end{aug}

\received{\smonth{3} \syear{2011}}
\revised{\smonth{10} \syear{2011}}

\begin{abstract}
Whole and targeted sequencing of human genomes is a promising,
increasingly feasible tool for discovering genetic contributions to risk
of complex diseases. A key step is calling an individual's genotype from
the multiple aligned short read sequences of his DNA, each of which is
subject to nucleotide read error. Current methods are designed to call
genotypes separately at each locus from the sequence data of unrelated
individuals. Here we propose likelihood-based methods that improve
calling accuracy by exploiting two features of sequence data. The first
is the linkage disequilibrium (LD) between nearby SNPs. The second is
the Mendelian pedigree information available when related individuals
are sequenced. In both cases the likelihood involves the probabilities
of read variant counts given genotypes, summed over the unobserved
genotypes. Parameters governing the prior genotype distribution and the
read error rates can be estimated either from the sequence data itself
or from external reference data. We use simulations and synthetic read
data based on the 1000 Genomes Project to evaluate the performance of
the proposed methods. An R-program to apply the methods to small
families is freely available at
\url{http://med.stanford.edu/epidemiology/PHGC/}.
\end{abstract}


\end{frontmatter}

\section{Introduction}\label{sec1}

\begin{figure}

\includegraphics{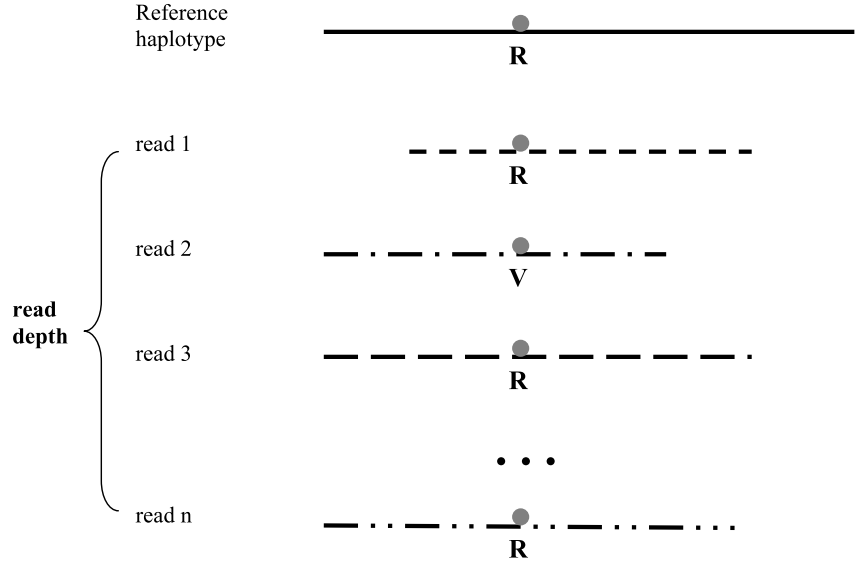}

  \caption{Alignment of n reads at a specific locus to a reference
haplotype, and calls of nucleotides ($R = {}$reference allele, $V = {}$variant)
at the locus. The $V$ at read 2 indicates a~SNP or a read error.}\label{fig1}
\end{figure}

The cost of DNA sequencing has decreased by orders of magnitude, and
further decreases will make it possible to sequence the entire human
genome in hundreds or thousands of individuals [Bentley et al.
(\citeyear{Ben08}),
Drma\-nac et al. (\citeyear{Drmetal10}), McKernan et al. (\citeyear{McK09})]. The resulting
comprehensive genomic analyses will provide powerful tools for
discovering the genetic variation underlying complex traits. Much of
this variation consists of single nucleotide polymorphisms (SNPs), which
occur with great frequency on the human genome. Each of our paired
chromosomes contains one of two nucleotides or alleles at each of its
SNP loci. Current sequencing technologies provide comprehensive
evaluation of an individual's alleles at all SNPs in a specific genomic
region by using his DNA to produce tens of millions of short nucleotide
sequences called \textit{reads} that range from 30 to 350 base pairs in
length. The nucleotide sequence of each read is then aligned to that of
a haploid reference genome. Each locus on the reference genome is thus
represented by a variable number n of reads called the \textit{read
depth}, and a count can be taken of the number of reads scored with the
variant (nonreference) nucleotide (Figure \ref{fig1}). In the absence of
sequencing error, the reads for an individual homozygous at the locus
would contain either all reference or all variant nucleotides. In
contrast, the reads for an individual heterozygous at the locus would
contain roughly half variants and half reference alleles, with binomial
variability due to the random sampling of his two homologous alleles. In
both instances, calling the correct genotype from his sequence reads is
complicated by errors in base calling and sequence alignment. Because of
the binomial variability across the reads of heterozygotes, such errors
are more difficult to detect and correct for heterozygous individuals
than for homozygotes. It is well established that the resulting
genotyping errors can lead to increased type I error and decreased power
in genetic studies [Gordon et al. (\citeyear{Gor02}), Clayton et al. (\citeyear{Cla05})].

Several methods have been developed to infer a genotype from the number
of variants among the aligned reads for an individual at a specific
locus [Li et al. (\citeyear{Li08}), Bansal et al. (\citeyear{Ban10}), Martin et al. (\citeyear{autokey11}), see
Nielsen et al. (\citeyear{Nie11}) for a review]. The approach of Li et al. (\citeyear{Li08})
calls the genotype for each individual separately, using fixed
prespecified values for genotype and nucleotide read error
probabilities. This approach has been criticized on the grounds that the
specified parameters may be inappropriate for any given individual
[Bansal et al. (\citeyear{Ban10}), Martin et al. (\citeyear{autokey11})]. Instead, Martin et al.
(\citeyear{autokey11}) use the sequence data in a sample of individuals from a given
racial/ethnic population to estimate allele frequencies and nucleotide
read error rates, and show that this strategy can improve genotype call
accuracy.

Both these approaches consider each locus separately, ignoring linkage
disequilibrium (LD) among nearby SNP loci, and both are designed for
application to unrelated individuals, not families. Two SNPs exhibit LD
when their minor alleles (those with population frequencies $\leq 0.5$)
are correlated in the chromosomes of the population. SNPs whose squared
correlation coefficients are close to one are said to be tightly linked
or in high LD. We shall show that for multiple SNPs in tight LD,
genotype calling accuracy can be improved by exploiting the genotype
constraints imposed by the LD. For example, if two tightly linked loci
have reads of different depths or reads with different error rates, the
genotypes at one locus can help determine the genotypes at the other. We
also shall show that the accuracy for sequenced family members can be
improved by exploiting the genotype constraints imposed by Mendelian
inheritance. The prospects for improvement are strengthened by the
success of methods for calling genotypes from the intensities produced
by SNP microarray chips: accuracy can be improved by modeling both SNP
LD [Yu et al. (\citeyear{Yu})] and Mendelian inheritance [Sabatti and Lange
(\citeyear{SabLan08}), Lin et al. (\citeyear{Linetal08})].

We begin by describing a likelihood-based algorithm for calling
genotypes at multiple loci in a region of high LD, either in sequenced
individuals or sequenced family members. We then apply the methods to
data from simulations and the 1000 Genomes Project. We conclude with a
brief discussion.

\section{Methods}\label{sec2}

We wish to call the genotypes of sequenced individuals in each of $I$
unrelated families for a set of $M$ SNPs in a given chromosomal region,
with unrelated individuals corresponding to `families' of size one. We
code an individual's genotype for a SNP according to the number $g = 0,
1$ or $2$ of minor SNP alleles he inherits from his parents, where, for
example, $g = 1$ indicates he received the minor SNP allele from one
parent but not the other. We wish to infer this genotype using the count
$y$ of variants observed in $n$ reads at the SNP, where for definiteness
we assume that the variant is the minor allele of the SNP. The number of
variant reads at a SNP locus depends on the individual's true genotype,
the number of reads (read depth), and the nucleotide error rate of the
reads at the locus.

Calling genotypes from such read data involves three steps: (1)
development of models for the distribution of unobserved genotypes (the
prior genotype distribution) and for the conditional distributions of
read variants given genotypes and read depth; (2) estimation of the
model parameters~$\bolds{\theta}$ by the method of maximum likelihood;
and (3) calling genotypes as the modes of the posterior distribution of
genotypes, with $\bolds{\theta}$ replaced by the maximum likelihood
estimate $\hat{\bolds{\theta}}$.

\textit{Step 1. Model development}. To describe the likelihood
of the observed read data for one family, let $\bolds{\theta} = (
\bolds{\pi} ,\bolds{\alpha} )$, where $\bolds{\pi}$ denotes the
parameters in the prior distribution of genotypes (or diploid
haplotypes) and $\bolds{\alpha}$ denotes the error parameters in the
conditional distribution of read variant counts given genotypes and read
depth. We assume negligible probability of recombination among the $M$
SNPs in the meioses connecting family members. That is, we assume the
region containing the $M$ SNPs is inherited as a single multi-allelic
locus with $H$ `alleles' labeled $1,2,\ldots,H$, where $H = 2^{M}$ is the
number of possible haplotypes in the population. Thus, each individual
carries one of the $H( H - 1 )/2$ possible haplotype pairs $( h,h' ),1
\le h \le h' \le H$, called diploid haplotypes. His genotype for SNP $m$
is $g_{m} = x_{mh} + x_{mh'}$, where $x_{mh} = 1$ if haplotype $h$
contains the minor allele at locus $m$, and zero otherwise. His
genotypes for the $M$ SNPs form a column vector $\mathbf{g} = (
g_{1},\ldots,g_{M} )^{T}$ of minor allele counts. Similarly, the
individual's read and variant data form column vectors $\mathbf{n} = (
n_{1},\ldots,n_{M} )^{T}$ and $\mathbf{y} = ( y_{1},\ldots,y_{M} )^{T}$ of
counts of reads and variants, respectively, at each the $M$ loci.

The model has the form
%
\begin{equation}
\label{eq1}
L( \bolds{\theta} ) = \Pr(
\mathbf{Y}|\mathbf{N},R;\bolds{\theta} ) = \sum_{\mathbf{G}}
\Pr( \mathbf{G}|R,\bolds{\pi} )\prod_{s = 1}^{S} \Pr(
\mathbf{y}_{s}|\mathbf{n}_{s},\mathbf{g}_{s};\bolds{\alpha} ) ,
\end{equation}
where $\mathbf{Y} =
(\mathbf{y}_{1},\mathbf{y}_{2},\ldots,\mathbf{y}_{S})$ and $\mathbf{N} =
(\mathbf{n}_{1},\mathbf{n}_{2},\ldots,\mathbf{n}_{S})$ are the $M \times S$
matrices of minor allele and read counts, respectively, for the $S$
sequenced family members. Also, the summation is taken over all possible
values for the $M \times S$ genotype matrix $\mathbf{G} =
(\mathbf{g}_{1},\mathbf{g}_{2},\ldots,\mathbf{g}_{S})$ that are consistent
with the members' relationship $R$. The prior genotype probabilities
$\Pr( \mathbf{G}|R;\bolds{\pi} )$ for the $S$ family members depend on
their relationship $R$ and on the parameters $\bolds{\pi}$ governing the
probabilities of the unrelated family founders' genotypes. To simplify
the notation, we model these probabilities assuming independence of
haplotypes within pedigree founders [i.e., Hardy--Weinberg (HW)
diploid haplotype frequencies] and across founders (i.e.,
random mating). However, departures from these assumptions can be
handled by more general modeling of the family genotype probabilities
$\Pr( \mathbf{G}|R;\bolds{\pi} )$ in (\ref{eq1}).

To model an individual's read variant counts at the $M$ loci, we assume
that, conditioned on his underlying genotype and total number of reads,
the read variant counts $y_{m}$ are independent and follow the binomial
model of Li et al. (\citeyear{Li08}), Kim et al. (\citeyear{Kim10}), and Martin et al. (\citeyear{autokey11}).
This model gives the probability of a variant count for an individual
with genotype $g_{m}$ at locus $m$ in terms of a nucleotide error rate
$\alpha_{m}$, defined as the probability that a read yields an incorrect
allele at locus~$m$. Specifically,
\begin{eqnarray}\label{eq2}
P(y_{m}|n_{m},g_{m} = 0;\alpha_{m}) &=& \pmatrix{ n_{m}
\vspace*{2pt}\cr
y_{m} }\alpha_{m}^{y_{m}}( 1 - \alpha_{m}
)^{n_{m} - y_{m}},
\nonumber\\
P(y_{m}|n_{m},g_{m} = 1;\alpha_{m}) &=& \pmatrix{ n_{m} \vspace*{2pt}\cr
y_{m}}\biggl( \frac{1}{2} \biggr)^{n_{m}},\\
P(y_{m}|n_{m},g_{m} = 2;\alpha_{m}) &= &\pmatrix{ n_{m}
\vspace*{2pt}\cr
y_{m}} ( 1 - \alpha_{m} )^{y_{m}}(
\alpha_{m} )^{n_{m} - y_{m}}.\nonumber
\end{eqnarray}
Like previous work [Li et al. (\citeyear{Li08}), Bansal et al. (\citeyear{Ban10}), Martin et
al. (\citeyear{autokey11})], this model assumes a single error probability for an
individual's reads at a SNP regardless of his actual genotype, an
assumption that can be relaxed by introducing allele-specific error
parameters. The model also assumes that read errors are independent both
within the reads at a single SNP and across the reads at different SNPs.
The latter assumption is tenable provided the SNPs of an inferred
diploid haplotype are at least half a kilobase apart and thus unlikely
to have overlapping reads.

\textit{Step 2. Likelihood maximization}. We
seek a value $\bolds{\hat{\theta}}$ to maximize the likelihood $L(
\bolds{\theta} ) = \prod_{i = 1}^{I} L_{i}( \bolds{\theta} )$ for all
the families, with $L_{i}( \bolds{\theta} )$ given by (\ref{eq1}). Finding
$\hat{\bolds{\theta}}$ can be facilitated in two ways. First, we can
compute the joint genotype probabilities of sequenced family members in
terms of their possible patterns of allele sharing due to its
inheritance from a common ancestor, called Identity-By-Descent (IBD)
sharing. That is, we can write $\Pr( \mathbf{G}|R;\bolds{\pi} ) =
\sum_{\phi} \Pr( \phi |R )\Pr( \mathbf{G}|\phi ;\bolds{\pi} )$, where
$\phi$ denotes an IBD configuration class for the members, and the
summation is taken over all classes consistent with the members'
relationship $R$ [Thompson (\citeyear{Tho74}), Whittemore and Halpern (\citeyear{WhiHal94})].
Second, we can use the EM algorithm [Dempster, Laird, and Rubin (\citeyear{DemLaiRub77})],
with the unobserved genotypes treated as missing data. The EM algorithm
is useful because, according to the model (1)--(2), the complete data
likelihood factors as a term involving the genotype parameters
$\bolds{\pi}$ times a term involving the nucleotide error rates
$\bolds{\alpha} = ( \alpha_{1},\ldots,\alpha_{M} )$. The \hyperref[app]{Appendix} contains
brief descriptions of these procedures.

\textit{Step 3. Genotype calling}. Finally,
we use the model (1)--(2) and its parameter estimate
$\hat{\bolds{\theta}}$ to call the genotypes for the $i${th} family as
the mode of the posterior distribution
\[
\Pr(
\mathbf{G}_{i}|\mathbf{N}_{i},\mathbf{Y}_{i};\hat{\bolds{\theta}}
) = \frac{\Pr ( \mathbf{G}_{i}|\hat{\bolds{\pi}} )\Pr
(
\mathbf{Y}_{i}|\mathbf{N}_{i},\mathbf{G}_{i};\hat{\bolds{\alpha}}
)}{\sum_{\mathbf{G}} \Pr ( \mathbf{G}|\bolds{\hat{\pi}}
)\Pr (
\mathbf{Y}_{i}|\mathbf{N}_{i},\mathbf{G};\hat{\bolds{\alpha}} )}
.
\]
We have implemented the above procedures in the R programs PedGC (uses
pedigree information), HapGC (uses LD information), and PedHapGC (uses
both), which are freely available at
\texttt{\href{http://med.stanford.edu/epidemiology/PHGC/}{http://med.stanford.edu/epidemiology/}
\href{http://med.stanford.edu/epidemiology/PHGC/}{PHGC/}}.

\textit{Special cases}. When genotypes are called separately
for each SNP and we assume HW genotype frequencies and random mating for
family founders, the prior genotype parameter $\bolds{\pi}$ equals the
vector of HW SNP minor allele frequencies (MAFs) at the SNPs. For this
case and for unrelated individuals, the model \mbox{(1)--(2)} agrees with that
used in the genotype calling method of Martin et al. (\citeyear{autokey11}), implemented
in the software package SeqEM.

For simultaneous calls at $M = 2$ SNPs, the $H = 2^{M} = 4$ possible
haplotypes have a multinomial population distribution with parameter
$\bolds{\pi} = ( \pi_{1},\pi_{2},\pi_{3} )$, where $\pi_{4} = 1 -
\sum_{m = 1}^{3} \pi_{m}$. Here we have ordered the four haplotypes as
follows: (1) $A_{1}A_{2}$; (2) $A_{1}B_{2}$; (3) $B_{1}A_{2}$; and
(4) $B_{1}B_{2}$, where $A_{m}$ and $B_{m}$ represent the minor and
major alleles, respectively, at SNP locus~$m$.

\section{Simulation results}\label{sec3}

We used simulations to evaluate the gains associated with exploiting LD
and pedigree information. We first computed error rates for genotypes
obtained separately at a single SNP but jointly for family members using
the model (1)--(2). We compared these error rates with those obtained
assuming genotype independence within families, as implemented in the
software package SeqEM [Martin et al. (\citeyear{autokey11})]. To do so, we generated
1000 data sets, each comprising 100 families containing one of three
sets of sequenced members: parent/offspring trios, sib pairs, and
first-cousin pairs. We generated genotypes using the IBD configuration
classes described in the \hyperref[app]{Appendix}, Part A, with various HW MAFs. Given genotypes,
we generated read depths and variant counts independently across
subjects. The read depths were taken as positive Poisson variables whose
means were $\mu = 10$ or $\mu = 30$, and the variant counts were
generated as binomial variables according to (\ref{eq2}), with read error rates
of 0.5\%, 5\%, and 10\%. In practice, read depths may exhibit
extra-Poisson variability across individuals and across SNPs within an
individual. However, this additional variability is unlikely to affect
the summary performance statistics reported here, since the calling
methods under consideration are all based on models that condition on
the observed read counts.

\begin{sidewaystable}
\tablewidth=\textwidth
\caption{Genotype error rates\protect\tabnoteref[a]{t1} (\%) for single SNP}
\label{tab1}
\begin{tabular*}{\textwidth}{@{\extracolsep{\fill}}ld{1.12}d{1.12}d{1.13}d{1.13}d{1.13}d{1.13}@{}}
\hline
& \multicolumn{6}{c@{}}{\textbf{Read error rate (\%)}}\\[-6pt]
& \multicolumn{6}{c@{}}{\hrulefill}\\
& \multicolumn{2}{c}{$\bolds{0.5}$} & \multicolumn{2}{c}{$\bolds{5.0}$} &
\multicolumn{2}{c@{}}{$\bolds{10.0}$}\\[-6pt]
& \multicolumn{2}{c}{\hrulefill} & \multicolumn{2}{c}{\hrulefill} & \multicolumn{2}{c@{}}{\hrulefill}\\
\multicolumn{1}{@{}l}{\textbf{MAF}\tabnoteref[b]{t2} \textbf{(\%)}} &
\multicolumn{1}{c}{\textbf{PedGC}\tabnoteref[c]{t3}} & \multicolumn{1}{c}{\textbf{SeqEM}\tabnoteref[d]{t4}} &
\multicolumn{1}{c}{\textbf{PedGC}} & \multicolumn{1}{c}{\textbf{SeqEM}} & \multicolumn{1}{c}{\textbf{PedGC}} & \multicolumn{1}{c@{}}{\textbf{SeqEM}}\\
\hline
 & \multicolumn{6}{c}{\textbf{Trios}}\\
 & \multicolumn{6}{c}{read depth${}={}$10}\\
\phantom{0}0.1 & 0.01\ (3.67/0.01) & 0.02\ (8.05/0.01) & 0.06\ (20.70/0.02) & 0.09\ (30.97/0.03) & 0.10\ (35.76/0.03) & 0.16\ (54.06/0.05)\\
\phantom{0}1.0 & 0.08\ (3.10/0.02) & 0.15\ (6.13/0.03) & 0.39\ (13.97/0.09) & 0.57\ (22.30/0.14) & 0.72\ (27.83/0.17) & 1.05\ (41.43/0.23)\\
10.0 & 0.51\ (2.07/0.17) & 0.81\ (3.13/0.30) & 2.01\ (7.32/0.85) & 2.86\ (10.06/1.28) & 4.02\ (14.86/1.63) & 5.37\ (18.97/2.37)\\[3pt]
 & \multicolumn{6}{c}{read depth${}={}$30}\\
\phantom{0}0.1 & 0.00\ (0.00/0.00) & 0.00\ (0.00/0.00) & 0.00\ (0.68/0.00) & 0.01\ (1.36/0.00) & 0.01\ (4.70/0.00) & 0.02\ (7.13/0.01)\\
\phantom{0}1.0 & 0.00\ (0.00/0.00) & 0.00\ (0.01/0.00) & 0.01\ (0.34/0.00) & 0.02\ (0.65/0.01) & 0.06\ (2.26/0.01) & 0.10\ (3.63/0.04)\\
10.0 & 0.00\ (0.01/0.00) & 0.00\ (0.01/0.00) & 0.05\ (0.20/0.02) & 0.09\ (0.32/0.03) & 0.30\ (0.99/0.15) & 0.46\ (1.48/0.24)\\[6pt]
 & \multicolumn{6}{c}{\textbf{Sib pairs}}\\
 & \multicolumn{6}{c}{read depth${}={}$10}\\
\phantom{0}0.1 & 0.01\ (5.99/0.01) & 0.02\ (8.71/0.01) & 0.06\ (21.05/0.02) & 0.09\ (28.57/0.03) & 0.13\ (43.23/0.05) & 0.18\ (58.62/0.07)\\
\phantom{0}1.0 & 0.10\ (3.89/0.03) & 0.14\ (5.43/0.03) & 0.46\ (16.88/0.13) & 0.60\ (22.70/0.16) & 0.84\ (30.80/0.23) & 1.08\ (42.17/0.25)\\
10.0 & 0.66\ (2.60/0.24) & 0.79\ (3.07/0.29) & 2.44\ (8.53/1.18) & 2.85\ (10.18/1.25) & 4.71\ (16.57/2.10) & 5.44\ (19.21/2.41)\\[3pt]
 & \multicolumn{6}{c}{read depth${}={}$30}\\
\phantom{0}0.1 & 0.00\ (0.00/0.00) & 0.00\ (0.00/0.00) & 0.00\ (0.26/0.00) & 0.00\ (1.05/0.00) & 0.01\ (5.97/0.01) & 0.02\ (7.27/0.01)\\
\phantom{0}1.0 & 0.00\ (0.00/0.00) & 0.00\ (0.00/0.00) & 0.01\ (0.46/0.00) & 0.01\ (0.64/0.01) & 0.08\ (2.64/0.03) & 0.12\ (4.00/0.04)\\
10.0 & 0.00\ (0.00/0.00) & 0.00\ (0.00/0.00) & 0.07\ (0.26/0.03) & 0.08\ (0.29/0.04) & 0.39\ (1.28/0.19) & 0.44\ (1.45/0.22)\\
\hline
\end{tabular*}
\end{sidewaystable}

\setcounter{table}{0}
\begin{sidewaystable}
\tablewidth=\textwidth
\caption{(Continued)}
\begin{tabular*}{\textwidth}{@{\extracolsep{\fill}}ld{1.12}d{1.13}d{1.13}d{1.13}d{1.13}d{1.13}@{}}
\hline
& \multicolumn{6}{c@{}}{\textbf{Read error rate (\%)}}\\[-6pt]
& \multicolumn{6}{c@{}}{\hrulefill}\\
& \multicolumn{2}{c}{$\bolds{0.5}$} & \multicolumn{2}{c}{$\bolds{5.0}$} &
\multicolumn{2}{c@{}}{$\bolds{10.0}$}\\[-6pt]
& \multicolumn{2}{c}{\hrulefill} & \multicolumn{2}{c}{\hrulefill} & \multicolumn{2}{c@{}}{\hrulefill}\\
\multicolumn{1}{@{}l}{\textbf{MAF}\tabnoteref[b]{t2} \textbf{(\%)}} &
\multicolumn{1}{c}{\textbf{PedGC}\tabnoteref[c]{t3}} & \multicolumn{1}{c}{\textbf{SeqEM}\tabnoteref[d]{t4}} &
\multicolumn{1}{c}{\textbf{PedGC}} & \multicolumn{1}{c}{\textbf{SeqEM}} & \multicolumn{1}{c}{\textbf{PedGC}} & \multicolumn{1}{c@{}}{\textbf{SeqEM}}\\
\hline
 & \multicolumn{6}{c}{\textbf{First-cousin pairs}}\\
 & \multicolumn{6}{c}{read depth${}={}$10}\\
\phantom{0}0.1 & 0.02\ (8.05/0.01) & 0.03\ (10.12/0.01) & 0.10\ (30.97/0.04) & 0.11\ (33.33/0.04) & 0.17\ (49.35/0.07) & 0.18\ (53.98/0.08)\\
\phantom{0}1.0 & 0.13\ (5.52/0.03) & 0.14\ (6.12/0.03) & 0.55\ (22.74/0.14) & 0.57\ (23.59/0.14) & 0.99\ (40.85/0.23) & 1.01\ (42.63/0.23)\\
10.0 & 0.80\ (3.09/0.30) & 0.81\ (3.36/0.34) & 2.79\ (10.16/1.21) & 2.84\ (10.95/1.37) & 5.37\ (18.59/2.70) & 5.43\ (20.31/2.73)\\[3pt]
 & \multicolumn{6}{c}{read depth${}={}$30}\\
\phantom{0}0.1 & 0.00\ (0.00/0.00) & 0.00\ (0.00/0.00) & 0.00\ (1.02/0.00) & 0.01\ (1.28/0.00) & 0.02\ (6.54/0.01) & 0.02\ (7.30/0.01)\\
\phantom{0}1.0 & 0.00\ (0.02/0.00) & 0.00\ (0.03/0.00) & 0.01\ (0.64/0.01) & 0.02\ (0.67/0.01) & 0.10\ (4.08/0.03) & 0.11\ (4.10/0.03)\\
10.0 & 0.01\ (0.02/0.00) & 0.01\ (0.02/0.00) & 0.09\ (0.33/0.05) & 0.09\ (0.35/0.05) & 0.44\ (1.41/0.24) & 0.45\ (1.55/0.25)\\
\hline
\end{tabular*}
\tabnotetext[a]{t1}{Percent incorrect genotype calls among sequenced members of
100 families of each type, averaged over 1000 replications. Numbers in
parenthesis give percent incorrect calls among heterozygote/homozygote
genotypes.}
\tabnotetext[b]{t2}{Minor allele frequency.}
\tabnotetext[c]{t3}{Uses individuals' familial relationships.}
\tabnotetext[d]{t4}{Treats individuals as unrelated.}
\end{sidewaystable}

Table \ref{tab1} shows the overall percentages of genotype errors when the data
were analyzed using PedGC and SeqEM. Also shown in parentheses are the
percentages of errors among individuals whose true genotypes at the SNP
are heterozygote and homozygote. Several observations are noteworthy.
First, as expected, accuracy increases with decreasing read error rates
and increasing read depth. Second, genotype errors are more common among
heterozygote genotypes than homozygous ones, consistent with the greater
difficulty in calling heterozygotes noted in the \hyperref[sec1]{Introduction}. Third,
the error rate among heterozygotes has relatively little impact on the
overall error rate, since genotypes that are heterozygous for rare
variants comprise a~small fraction of the population. Nevertheless, as
noted in the \hyperref[sec5]{Discussion}, both types of error can impair one's
ability to distinguish heterozygote from normal homozygote individuals
with the accuracy needed for good power in association studies. Fourth,
larger MAFs are associated with increased accuracy among heterozygotes
but decreased accuracy among homozygotes, and thus decreased overall
accuracy. Finally, PedGC consistently improves the error rate among both
heterozygotes and homozygotes, and thus consistently improves the
overall error rate.

{\renewcommand{\thetable}{\arabic{table}A}
\begin{sidewaystable}
\tablewidth=\textwidth
\caption{Error rates\protect\tabnoteref[a]{t21} (\%) for genotypes of SNP pairs in unrelated individuals}
\label{tab2a}
\begin{tabular*}{\textwidth}{@{\extracolsep{\fill}}lccccd{1.13}d{1.13}d{1.13}d{1.13}@{}}
\hline
&&&& & \multicolumn{4}{c@{}}{\textbf{Unrelated individuals}}\\[-6pt]
&&&& & \multicolumn{4}{c@{}}{\hrulefill}\\
\multicolumn{2}{c}{\textbf{MAF}\tabnoteref[b]{t22} (\%)} & \multicolumn{2}{c}{\textbf{Read error rate (\%)}} &  & \multicolumn{2}{c}{\textbf{SNP1}} &
\multicolumn{2}{c@{}}{\textbf{SNP2}}\\[-6pt]
\multicolumn{2}{@{}l}{\hrulefill} & \multicolumn{2}{c}{\hrulefill} &  & \multicolumn{2}{c}{\hrulefill} &
\multicolumn{2}{c@{}}{\hrulefill}\\
\textbf{SNP1} & \textbf{SNP2} & \textbf{SNP1} & \textbf{SNP2} & \multicolumn{1}{c}{$\bolds{r^2}$} &
\multicolumn{1}{c}{\textbf{HapGC}\tabnoteref[c]{t23}} & \multicolumn{1}{c}{\textbf{SeqEM}\tabnoteref[d]{t24}} &
\multicolumn{1}{c}{\textbf{HapGC}} & \multicolumn{1}{c@{}}{\textbf{SeqEM}}\\
\hline
 &  &  &  &  & \multicolumn{4}{c@{}}{\textbf{read depth${}={}$10}} \\
1.0 & 1.0 & 5.0 & 5.0 & 0.9 & 0.19\ (7.11/0.05) & 0.60\ (22.61/0.15) & 0.18\ (7.23/0.04) & 0.60\ (22.41/0.15)\\
 &  &  &  & 0.8 & 0.24\ (8.58/0.07) & 0.62\ (24.10/0.15) & 0.26\ (9.98/0.07) & 0.63\ (24.31/0.16)\\
 &  &  &  & 0.6 & 0.30\ (11.45/0.08) & 0.61\ (22.25/0.17) & 0.29\ (11.00/0.08) & 0.60
\ (21.48/0.18)\\[3pt]
 &  & 1.0 & 5.0 & 0.9 & 0.07\ (2.98/0.01) & 0.20\ (7.49/0.05) & 0.14\ (5.21/0.04) & 0.59\ (22.80/0.14)\\
 &  &  &  & 0.8 & 0.09\ (3.68/0.02) & 0.21\ (7.76/0.05) & 0.20\ (7.19/0.06) & 0.63\ (23.48/0.16)\\
 &  &  &  & 0.6 & 0.11\ (4.54/0.02) & 0.20\ (7.46/0.05) & 0.29\ (11.01/0.07) & 0.61
\ (24.00/0.14)\\[6pt]
 &  &  &  &  & \multicolumn{4}{c@{}}{\textbf{read depth${}={}$30}}\\
1.0 & 1.0 & 5.0 & 5.0 & 0.9 & 0.01\ (0.20/0.00) & 0.02\ (0.90/0.01) & 0.00\ (0.17/0.00) & 0.01\ (0.57/0.01)\\
 &  &  &  & 0.8 & 0.01\ (0.15/0.00) & 0.02\ (0.88/0.01) & 0.00\ (0.15/0.00) & 0.02\ (0.95/0.01)\\
 &  &  &  & 0.6 & 0.01\ (0.33/0.00) & 0.02\ (0.66/0.01) & 0.01\ (0.32/0.00) & 0.02\
 (0.72/0.00)\\[3pt]
 &  & 1.0 & 5.0 & 0.9 & 0.00\ (0.02/0.00) & 0.00\ (0.05/0.00) & 0.00\ (0.20/0.00) & 0.02\ (0.83/0.01)\\
 &  &  &  & 0.8 & 0.00\ (0.00/0.00) & 0.00\ (0.00/0.00) & 0.01\ (0.32/0.00) & 0.02\ (0.80/0.01)\\
 &  &  &  & 0.6 & 0.00\ (0.02/0.00) & 0.00\ (0.02/0.00) & 0.01\ (0.44/0.00) & 0.02\ (0.96/0.01)\\
\hline
\end{tabular*}
\end{sidewaystable}}

{\renewcommand{\thetable}{\arabic{table}B}
\setcounter{table}{1}
\begin{sidewaystable}
\tabcolsep=0pt
\tablewidth=\textwidth
\caption{Error rates\protect\tabnoteref[a]{t21} (\%) for genotypes of SNP pairs in first-cousin pairs}
\label{tab2b}
\begin{tabular*}{\textwidth}{@{\extracolsep{\fill}}lccccd{1.13}d{1.13}d{1.13}d{1.13}d{1.13}d{1.13}@{}}
\hline
&&&& & \multicolumn{6}{c@{}}{\textbf{First-cousin pairs}}\\[-6pt]
&&&& & \multicolumn{6}{c@{}}{\hrulefill}\\
\multicolumn{2}{c}{\textbf{MAF}\tabnoteref[b]{t22} (\%)} & \multicolumn{2}{c}{\multirow{2}{45pt}[10pt]{\centering\textbf{Read error rate (\%)}}} &  & \multicolumn{3}{c}{\textbf{SNP1}} &
\multicolumn{3}{c@{}}{\textbf{SNP2}}\\[-6pt]
\multicolumn{2}{@{}l}{\hrulefill} & \multicolumn{2}{l}{\hrulefill} &  & \multicolumn{3}{c}{\hrulefill} &
\multicolumn{3}{c@{}}{\hrulefill}\\
\textbf{SNP1} & \textbf{SNP2} & \textbf{SNP1} & \textbf{SNP2} & \multicolumn{1}{c}{$\bolds{r^2}$} &
\multicolumn{1}{c}{\textbf{PedHapGC}\tabnoteref[e]{t25}} & \multicolumn{1}{c}{\textbf{PedGC}\tabnoteref[f]{t26}} &
\multicolumn{1}{c}{\textbf{SeqEM}} & \multicolumn{1}{c@{}}{\textbf{PedHapGC}}& \multicolumn{1}{c}{\textbf{PedGC}} & \multicolumn{1}{c@{}}{\textbf{SeqEM}}\\
\hline
 &  &  &  &  & \multicolumn{6}{c}{\textbf{read depth${}={}$10}}\\
1.0 & 1.0 & 5.0 & 5.0 & 0.9 & 0.16\ (6.26/0.04) & 0.55\ (20.73/0.14) & 0.59\ (22.82/0.14) & 0.19\ (6.92/0.05) & 0.54\ (20.26/0.14) & 0.58\ (22.20/0.14)\\
 &  &  &  & 0.8 & 0.21\ (8.28/0.05) & 0.56\ (21.40/0.15) & 0.61\ (23.82/0.15) & 0.23\ (8.75/0.06) & 0.56\ (21.72/0.14) & 0.61\ (24.20/0.14)\\
 &  &  &  & 0.6 & 0.29\ (11.57/0.07) & 0.57\ (21.76/0.14) & 0.61\ (23.99/0.14) & 0.30\ (11.51/0.07) & 0.58\ (21.86/0.15) & 0.62
\ (24.12/0.15)\\[3pt]
 &  & 1.0 & 5.0 & 0.9 & 0.07\ (2.85/0.01) & 0.20\ (7.87/0.05) & 0.22\ (8.77/0.05) & 0.15\ (5.32/0.04) & 0.59\ (20.98/0.17) & 0.63\ (23.43/0.16)\\
 &  &  &  & 0.8 & 0.09\ (3.63/0.02) & 0.20\ (7.69/0.05) & 0.21\ (8.37/0.05) & 0.18\ (6.51/0.06) & 0.57\ (21.55/0.15) & 0.62\ (23.99/0.15)\\
 &  &  &  & 0.6 & 0.11\ (4.42/0.02) & 0.20\ (7.56/0.05) & 0.21\ (8.39/0.05) & 0.29\ (10.35/0.08) & 0.58\ (22.02/0.15) & 0.62
\ (24.31/0.15)\\[6pt]
 &  &  &  &  & \multicolumn{6}{c}{\textbf{read depth${}={}$30}}\\
1.0 & 1.0 & 5.0 & 5.0 & 0.9 & 0.01\ (0.20/0.001) & 0.01\ (0.57/0.01) & 0.02\ (0.67/0.01) & 0.00\ (0.10/0.00) & 0.02\ (0.68/0.01) & 0.02\ (0.78/0.01)\\
 &  &  &  & 0.8 & 0.01\ (0.24/0.003) & 0.02\ (0.74/0.01) & 0.02\ (0.79/0.01) & 0.00\ (0.17/0.00) & 0.02\ (0.57/0.01) & 0.02\ (0.67/0.01)\\
 &  &  &  & 0.6 & 0.01\ (0.24/0.002) & 0.02\ (0.52/0.01) & 0.02\ (0.66/0.01) & 0.01\ (0.39/0.00) & 0.02\ (0.66/0.01) & 0.02
\ (0.74/0.01)\\[3pt]
 &  & 1.0 & 5.0 & 0.9 & 0.00\ (0.00/0.00) & 0.00\ (0.00/0.00) & 0.00\ (0.00/0.00) & 0.01\ (0.14/0.00) & 0.02\ (0.59/0.01) & 0.02\ (0.71/0.01)\\
 &  &  &  & 0.8 & 0.00\ (0.00/0.00) & 0.00\ (0.02/0.00) & 0.00\ (0.05/0.00) & 0.01\ (0.36/0.00) & 0.03\ (1.11/0.01) & 0.03\ (1.19/0.01)\\
 &  &  &  & 0.6 & 0.00\ (0.00/0.00) & 0.00\ (0.05/0.00) & 0.00\ (0.07/0.00) & 0.01\ (0.50/0.00) & 0.02\ (0.78/0.01) & 0.02\ (0.81/0.01)\\
\hline
\end{tabular*}
\tabnotetext[a]{t21}{Percent incorrect genotype calls among 100 families, averaged
over 1000 replications.}
\tabnotetext[b]{t22}{Minor allele frequency; each SNP has MAF${} = {}$1\%.}
\tabnotetext[c]{t23}{Uses SNP LD.}
\tabnotetext[d]{t24}{Treats locus genotypes as independent and family members as
unrelated.}
\tabnotetext[e]{t25}{Uses LD and familial relationships.}
\tabnotetext[f]{t26}{Uses familial relationships.}
\end{sidewaystable}}

To evaluate gains in accuracy from incorporating LD information, we also
generated read data for $M=2$ loci in high LD, and specified the HW
haplotype frequency parameter $\bolds{\pi} = ( \pi_{1},\pi_{2},\pi_{3}
)$ in terms of the SNP MAFs $p_{1}, p_{2}$ and their correlation
coefficient $r$. In each of 1000 replications, we generated
two-locus diploid haplotypes for 100 unrelated individuals and 100
first-cousin pairs, as described in the \hyperref[app]{Appendix}, Part A. Read depths were taken
to be positive Poisson variables distributed independently across the
two loci and across individuals, and variant counts and read errors were
generated independently, as described above. Read error parameters
$\alpha_{m}$ were allowed to differ at the two loci. Tables \ref{tab2a} and~\ref{tab2b} give
results for the following: (A) unrelated individuals and (B)
first-cousins. The table shows that pedigree and LD information improves
heterozygote- and homozygote-specific genotype call accuracy for linked
loci in both unrelated individuals and families. Accuracy gains are
particularly large for SNPs that are tightly linked to another SNP with
lower read error rates. Accuracy gains diminish with decreasing LD
between the loci, and with increasing read depth.

The data for Tables \ref{tab1}, \ref{tab2a} and \ref{tab2b} were generated assuming HW frequencies and
random mating in family founders. To evaluate the impact on genotype
calling of departures from these assumptions, we also evaluated the
performance of the calling methods as applied to data generated with
departures from these assumptions. Specifically, we generated pedigree
founder genotypes with excess homozygity as well as excess
heterozygosity, and with various levels of assortative mating. We found
little change in the accuracy gains for HapGC and PedGC relative to
SeqEM (data not shown).

\section{Application to 1000 genomes data}\label{sec4}

\begin{figure}

\includegraphics{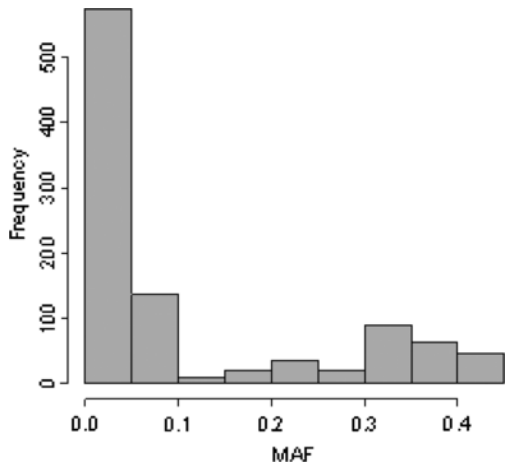}

  \caption{Histogram of minor allele frequencies of the 1000 SNPs
identified in a 274~kb region of chromosome 21 among 283 Caucasian
participants in the 1000 Genomes Project.}\label{fig2}
\end{figure}

The 1000 genomes project is an international research initiative to
catalogue human genetic variants by sequencing at least one thousand
subjects representing different racial/ethnic groups. In a publicly
available data from a pilot phase, exomic sequencing has identified some
15~million variants, more than half of which were previously unknown
[The 1000 Genomes Project Consortium (\citeyear{Con10})]. To illustrate the LD- and
pedigree-based methods as applied to the haplotypes of family members
from a Caucasian population, we focused on a randomly chosen 274~kb
region from position 17,345,389 to position 17,619,118 on chromosome 21,
for which 1000 SNPs were found among the 283 Caucasian subjects. We
chose a region of this length to allow enough SNPs to evaluate the
methods across a broad range of SNP MAFs and pairwise SNP correlation
coefficients. We chose a region containing 1000 SNPs to allow evaluation
of the methods over a broad range of variant frequencies and pairs of
SNP correlation coefficients. Figure \ref{fig2} shows the distribution of MAFs
for these 1000 SNPs, as obtained from the set of $2\times 283 = 586$
phased Caucasian haplotypes.

We used these 586 haplotypes to generate diploid hapotypes for 100
families containing one of three sets of sequenced members:
parent--offspring trios, sib pairs, and first-cousin pairs. To do so, we
randomly and independently sampled with replacement from the 583
haplotypes to obtain diploid haplotypes for family members, using the
distribution of IBD configuration classes specific for their genetic
relationships (see the \hyperref[app]{Appendix}, Part A for details). Then, given an individual's
diploid haplotype, we generated read depths and read errors
independently across the 1000 SNPs. Read depths were generated as
independent positive Poisson variables. Read errors were generated
according to the model (2), with SNP-specific read error rates
$\alpha_{m}$ obtained by sampling from the uniform distribution on the
interval [0.001, 0.1].

Data analysis using HapGC or PedHapGC involves calling the genotypes for
all SNPs in a region in three steps. In step 1, PedGC is used to call
separately the genotypes at each SNP. In step 2, these genotypes are
used to estimate correlation coefficients for all pairs of SNPs in the
region. In step 3, SNP-specific genotypes are again called
simultaneously with a second SNP chosen to have low estimated read error
rate and high correlation coefficient with the targeted SNP. For step 1,
we used PedGC to call genotypes separately for each of the 1000 SNPs in
the chosen region, using $I = 5, 25, 50$, and all 100 families, and then
implemented steps~2 and 3. Table \ref{tab3} shows genotype error rates averaged
over the 1000 SNPs, using various read depths and numbers of families.
PedGC produced lower genotyping error than SeqEM for both heterozygotes
and homozygotes. The gain in accuracy increased with increasing numbers
of families, up to $I=25$ families, with little increase thereafter.
These increases reflect increased precision of estimates for the
parameters $\pi$ governing joint family genotypes. As expected, overall
genotyping errors declined sharply with increasing read depth. For $n =
30$ reads, the genotyping error rate was one or two orders of magnitude
lower than the rate for $n=5$ reads.

\begin{sidewaystable}
\tablewidth=\textwidth
\tabcolsep=0pt
\caption{Genotype error rates\protect\tabnoteref[a]{t31} (\%) for 1000 SNPs\protect\tabnoteref[b]{t32}
called separately, using 1000 genomes data}
\label{tab3}
{\fontsize{8.3}{10.3}\selectfont{
\begin{tabular*}{\textwidth}{@{\extracolsep{\fill}}ld{1.14}d{1.13}d{1.13}d{1.13}d{1.13}d{1.13}d{1.13}d{1.13}@{}}
\hline
 & \multicolumn{8}{c@{}}{\textbf{Number $\bolds{I}$ of families\tabnoteref[c]{t33}}}\\[-6pt]
 & \multicolumn{8}{c@{}}{\hrulefill}\\
 & \multicolumn{2}{c}{\textbf{5}} & \multicolumn{2}{c}{\textbf{25}} & \multicolumn{2}{c}{\textbf{50}} &
 \multicolumn{2}{c@{}}{\textbf{100}}\\[-6pt]
  & \multicolumn{2}{c}{\hrulefill} & \multicolumn{2}{c}{\hrulefill} & \multicolumn{2}{c}{\hrulefill} & \multicolumn{2}{c@{}}{\hrulefill}\\
 \multicolumn{1}{@{}l}{\multirow{2}{30pt}[10pt]{\textbf{Read depth}}} & \multicolumn{1}{c}{\textbf{PedGC}} & \multicolumn{1}{c}{\textbf{SeqEM}} & \multicolumn{1}{c}{\textbf{PedGC}} & \multicolumn{1}{c}{\textbf{SeqEM}} &
 \multicolumn{1}{c}{\textbf{PedGC}} & \multicolumn{1}{c}{\textbf{SeqEM}} & \multicolumn{1}{c}{\textbf{PedGC}} & \multicolumn{1}{c@{}}{\textbf{SeqEM}}\\
 \hline
& \multicolumn{8}{c}{\textbf{Parent--offspring trios}}\\
\phantom{0}5 & 5.61\ (27.48/2.04) & 7.58\ (34.72/2.96) & 4.73\ (20.61/2.06) & 5.82\ (24.28/2.64) & 4.44\ (19.42/1.87) & 5.42\ (22.57/2.52) & 4.40\ (19.23/1.87) & 5.35\ (22.61/2.44)\\
10 & 1.98\ (9.46/0.99) & 2.64\ (13.85/1.25) & 1.60\ (6.57/0.82) & 2.34\ (8.81/1.17) & 1.63\ (6.59/0.75) & 2.19\ ( 8.88/1.07) & 1.58\ (6.79/0.71) & 2.13\ (9.06/1.00)\\
30 & 0.09\ (0.37/0.06) & 0.12\ (0.42/0.06) & 0.07\ (0.18/0.02) & 0.12\ ( 0.34/0.04) & 0.07\ (0.20/0.03) & 0.11\ (0.35/0.04) & 0.06\ (0.24/0.03) &
0.10\ (0.37/0.04)\\[3pt]
 & \multicolumn{8}{c}{\textbf{Sib pairs}}\\
\phantom{0}5 & 6.36\ (28.14//2.67) & 7.38\ (34.64/2.77) & 5.16\ (21.62/2.25) & 5.95\ (25.45/2.51) & 5.07\ (20.42/2.38) & 5.55\ (24.06/2.43) & 4.95\ (19.90/2.44) & 5.35\ (23.04/2.49)\\
10 & 2.42\ (10.99/0.97) & 3.16\ (15.00/1.15) & 2.02\ (8.18/0.93) & 2.28\ (8.62/1.16) & 1.93\ (7.60/0.95) & 2.16\ (8.57/1.05) & 1.89\ (7.64/0.93) & 2.10\ (8.78/0.98)\\
30 & 0.10\ (0.34/0.05) & 0.18\ (0.62/0.10) & 0.08\ (0.32/0.04) & 0.09\ (0.36/0.04) & 0.08\ (0.31/0.04) & 0.09\ (3.54/0.05) & 0.07\ (0.30/0.04) & 0.09\ (0.35/0.04)\\[3pt]
 & \multicolumn{8}{c}{\textbf{First-cousin pairs}}\\
\phantom{0}5 & 6.44\ (33.32/2.20) & 6.87\ (34.18/2.54) & 5.42\ (23.56/2.17) & 5.98\ (24.97/2.57) & 5.14\ (22.20/2.15) & 5.59\ (23.20/2.50) & 5.02\ (21.40/2.26) & 5.37\ (22.62/2.46)\\
10 & 2.48\ (12.51/0.89) & 2.80\ (13.68/1.07) & 2.10\ (9.06/0.85) & 2.26\ (9.08/1.03) & 2.01\ (8.63/0.85) & 2.20\ (9.00/1.00) & 1.98\ (8.46/0.89) & 2.10\ (8.84/0.97)\\
30 & 0.12\ (0.43/0.06) & 0.12\ (0.43/0.07) & 0.12\ (0.47/0.05) & 0.13\ (0.48/0.06) & 0.09\ (0.37/0.04) & 0.10\ (0.37/0.05) & 0.10\ (0.40/0.04) & 0.11\ (0.43/0.05)\\
\hline
\end{tabular*}}}
\tabnotetext[a]{t31}{Percent incorrect genotype calls among all \textit{1000SI}
calls for 1000 SNPs in $S$ sequenced members of $I$ families; read error
rates randomly chosen from [0.001, 0.1]. Numbers in parenthesis give
percent incorrect calls among heterozygotes/homozygotes.}
\tabnotetext[b]{t32}{SNPs lie in a 274~kb region of chromosome 21.}
\tabnotetext[c]{t33}{See the \hyperref[app]{Appendix}, Part A for description of how diploid haplotypes of
sequenced family members were obtained.}
\end{sidewaystable}

To evaluate the gains from exploiting LD, we also implemented steps 2
and 3 of the preceding paragraph. Specifically, we used the family
genotype calls for all SNPs in the 274 kb region (Table \ref{tab3}) to calculate
correlation coefficients for each of the 499,500 pairs of SNPs. We then
selected for each SNP another maximally correlated SNP, and called the
pair simultaneously. When based on the read data and 2-locus diploid
haplotypes (called from step 1) of the cousin pairs of Table \ref{tab3}, the
PedHapGC calls had error rates of 2.80\%, 0.80\%, and 0.03\% for read
depths of 5, 10, and 30, respectively. In comparison to the single locus
calls of PedGC, the genotype error rates were reduced by 44\%, 60\%, and
70\%, respectively. Compared to SeqEM, the reductions were 48\%, 62\%,
and 73\%, respectively. These findings support the simulation results,
suggesting that simultaneous exploitation of LD and pedigree
relationships can have a significant impact on calling accuracy.

The previous paragraphs describe the simultaneous call of genotypes for
each of a pair of SNPs in LD. In some applications, it may be more
appropriate to call an individual's entire diploid haplotype for the
SNPs of interest. To evaluate such haplotype calling, we attempted to
call the diploid haplotypes of the 100 first-cousin pairs created for
Table \ref{tab3}. Specifically, we focused on their diploid 3-locus haplotypes
for a specific set of three SNPs in the 274 kb region having MAFs of
28\%, 31\%, and 12\%, and pairwise correlation coefficients of 0.97,
0.94, and 0.92. There were six distinct haplotypes among the 283
Caucasian subjects. We generated 1000 replications of read data for each
of the 100 pairs of first-cousins, with nucleotide read error rates
again randomly sampled from a uniform distribution on the interval
[0.001, 0.1]. The diploid haplotype error rates using PedHapGC were
2.0\%, 0.4\%, and 0.02\% for read depths of 5, 10, and 30, respectively,
lower than the SNP-specific genotype error rates for 100 first-cousin
pairs shown in Table \ref{tab3}. These results suggest that diploid haplotypes
for three SNPs can be called accurately. However, the computational
burden involved in accommodating many possible diploid haplotypes
precludes using LD information for more than a few SNPs.

When available, the haplotypes of a reference panel from the same
population as that under study can be used in HapGC to specify the
parameters~$\bolds{\pi}$ governing the diploid haplotypes probabilities,
rather than estimating them from the sequenced data at hand. Simulations
suggest that specification of $\bolds{\pi}$ yields reasonable results
even when the specified values deviate from their actual values in the
population of interest. For example, when inferring haplotypes from 100
pairs of first-cousins, incorrect assignment of equal HW frequencies to
the haplotypes gave mean haplotype errors of 3.4\%, 1.1\%, and 0.1\% for
read depths of 5, 10, and 30, respectively. Comparison to the results of
the previous paragraph suggests that the loss in accuracy improvement is
not substantial.

\begin{table}
\tabcolsep=0pt
\caption{Genotype error rates\protect\tabnoteref[a]{t41} (\%) for sib pairs with
and without parental genotyping\protect\tabnoteref[b]{t42}}
\label{tab4}
\begin{tabular*}{\textwidth}{@{\extracolsep{\fill}}ld{1.13}d{1.13}d{1.13}d{1.13}@{}}
\hline
& \multicolumn{4}{c@{}}{\textbf{MAF}\tabnoteref[c]{t43}
 \textbf{(\%)}}\\[-6pt]
 & \multicolumn{4}{c@{}}{\hrulefill}\\
\multicolumn{1}{@{}l}{\centering\multirow{2}{45pt}[10pt]{\textbf{Read error rate (\%)}}} & \multicolumn{1}{c}{\textbf{0.1}} & \multicolumn{1}{c}{\textbf{1.0}} & \multicolumn{1}{c}{\textbf{10.0}} & \multicolumn{1}{c}{\textbf{20.0}}\\
\hline
 & \multicolumn{4}{c}{\textbf{Sibs only (10 reads per person)}}\\
1.0 & 0.00\ (68.83/0.00) & 0.66\ (31.72/0.01) & 1.58\ (5.48/0.72) & 2.34\ (3.19/1.93)\\
5.0 & 0.12\ (67.97/0.01) & 1.02\ (52.26/0.02) & 3.73\ (16.62/0.86) & 5.18\
(8.99/3.38)\\[3pt]
 & \multicolumn{4}{c}{\textbf{Sibs and parents (5 reads per person)}}\\
1.0 & 0.00\ (43.82/0.00) & 0.62\ (31.36/0.01) & 3.65\ (15.37/1.12) & 5.75\ (10.50/3.51)\\
5.0 & 0.14\ (77.77/0.00) & 1.19\ (57.91/0.02) & 6.87\ (31.57/1.45) & 9.84\ (19.31/5.48)\\
\hline
\end{tabular*}
\tabnotetext[a]{t41}{Percent incorrect genotype calls; read error rates randomly
chosen from [0.001, 0.1]. Numbers in parenthesis give percent incorrect
calls among heterozygotes/homozygotes.}
\tabnotetext[b]{t42}{For a given total number of reads.}
\tabnotetext[c]{t43}{Minor allele frequency.}
\end{table}

Finally, we examined the trade-off between providing more reads to key
individuals versus spreading fewer reads across the key
individuals and their relatives. This trade-off was motivated by the
results in Table \ref{tab3} showing that read depth is an important determinant
of genotype call accuracy, but that the number of sequenced families and
individuals plays a lesser role. When multiple family members can be
sequenced and the budget accommodates a fixed total number of reads, a
practical design question concerns the relative merits of sequencing
fewer family members with greater read depth vs sequencing more members
with lower depth. We addressed this question in the context of
genotyping sib pairs at a single locus. That is, we compared the
genotype accuracy associated with sequencing 50 sib pairs with 10 reads
per person (1000 reads in total) to that associated with sequencing 50
sibs and their parents, with 5 reads per person (also 1000 reads in
total). Table \ref{tab4} shows the genotype error rates for SNPs with MAFs
ranging from 0.001 to 0.2. For small MAFs, both strategies yielded
similar genotyping accuracy. But for larger MAFs, sequencing fewer
members with greater depth performed better. These results suggest that
little is gained from genotyping additional family members unless their
genotypes contribute independently to the study objectives.

\section{Discussion}\label{sec5}

We have presented a likelihood-based approach to calling genotypes and
haplotypes of sequenced family members. Simulations and application to
1000 Genomes data show that LD and pedigree information can be used to
improve the accuracy of called genotypes for individuals who are
heterozygous for rare variants, as well as for homozygous individuals,
who typically constitute the majority of subjects. However, there are
limits to the gains achieved using pedigree information: for example,
the simulations suggest that, for a given total number of reads, greater
calling accuracy is achieved by allocating all reads to those
individuals whose phenotypes contribute to the study goals than by
allocating some reads to additional ancillary family members. As
expected from the greater difficulty of calling heterozygotes than
homozygotes, error rates are considerably higher in heterozygotes than
homozygotes. But because most individuals are homozygous for the normal
SNP allele, genotype errors in these individuals dominate the overall
error rate.

A few notes about the strengths and limitations of the proposed methods
are warranted. The methods do not require external specification of
genotype frequencies, LD measures, or nucleotide read error
probabilities. In particular, HapGC and PedHapGC can be implemented with
the sequence data at hand without data from an external reference panel
as needed by hidden Markov LD models. However, this flexibility comes at
a price: although the methods put no constraints on the number of SNPs
called jointly, in practice, the number of haplotypes increases
exponentially with the number of SNPs in LD, and the computation
involved in parameter estimation becomes demanding for more than a few
SNPs. We suggest using the LD of two or three SNPs, as a trade-off
between computational efficiency and accuracy gain. The alternative is
to use externally-derived haplotype frequencies; limited simulations
(data not shown) suggest that this can be advantageous provided the
haplotype frequencies do not deviate substantially from those in the
population under study.

The model presented in equations (\ref{eq1}) and (\ref{eq2}) involves several
assumptions, including HW genotype frequencies and random mating for
family founders, and symmetric read error probabilities. None of these
assumptions is central to the model; they can be relaxed by including
additional parameters. Typically the number of families or unrelated
individuals sequenced is large, so estimates of these additional
parameters would be stable. However, while calling SNP genotypes in
targeted regions is apt to be feasible, the computational burden could
be limiting for whole genome sequencing of many individuals. Thus, it is
encouraging that the simulations showed gains in calling accuracy for
the proposed methods even in the presence of moderate departures from
the assumptions. An advantage of the current approach is that it is not
necessary to distinguish misalignment errors from erroneous nucleotide
calls, as the nucleotide read error probabilities $\alpha_{m}$ include
both sources of error. Thus, the assumption of independence of read
errors across loci is reasonable provided the loci are more than a few
hundred kilobases apart.\looseness=-1

While the simulations presented here show clear accuracy gains from
using LD and pedigree information, more complete evidence would derive
from actual rather than simulated read data for subjects whose DNA had
also been sequenced using a costly and highly accurate method (the gold
standard). Then any problems and trouble spots associated with, say,
heterogeneity across regions in depth coverage and error rate could be
examined empirically. However, we do not expect such heterogeneity
issues to affect the relative performance of the methods considered,
because they all allow region-specific estimation of error rates. If
there are computational constraints on the number of error rates in need
of estimation, it may prove useful to regress them on characteristics
predictive of accuracy, such as deviations from HW frequencies, aberrant
LD patterns, or low read depth. These issues will be examined in future
work.

In conclusion, even when nucleotide read error rates are low, the
genotype-calling improvements obtained using LD and pedigree information
can be important determinants of statistical power to detect
associations between disease and very rare alleles with only moderate
effect sizes using thousands of study subjects. For example, the impact
of misclassification error on the power of a test for variant-disease
association in a case--control study can be inferred from the seminal
work of Bross (\citeyear{Bro54}). As applied to a case--control study of carriage of
a rare variant, the author describes estimation bias and power loss when
cases and controls have the same probabilities of missing a variant
carrier, and the same probabilities of falsely detecting one. Under
these nondifferential error assumptions (which seem reasonable for the
current sequencing problem), Bross (\citeyear{Bro54}) provides expressions for the
proportional increase in sample size needed to achieve the same power
the study would provide in the absence of measurement error. His
calculations show that incorrectly detecting a variant carrier has more
serious adverse consequences for power than does missing such an
individual. Nevertheless, when the null hypothesis is rejected, both
types of errors can lead to serious bias in estimates of variant effect
size.

\begin{appendix}\label{app}
\section*{Appendix}

A. \textit{Using IBD relationships to compute family genotype
probabilities and generate family genotypes.} A set of alleles of family
members at a given locus is \textit{Identical-by-Descent (IBD)} if all
the alleles are inherited from a common ancestor.\vadjust{\goodbreak} The set is
\textit{IBD-distinct} if no two alleles are IBD. The joint genotype
probabilities for any given set S of pedigree members can be computed
using the IBD configuration classes (ICCs) described by Thompson (\citeyear{Tho74}),
or the inheritance vectors described by Kruglyak et al. (\citeyear{Kruetal96}). Loosely
speaking, an ICC specifies which subsets of the members' 2S alleles are
IBD. For example, the set of six alleles of a noninbred parent/offspring
trio has just one ICC, that in which the four parental alleles are
IBD-distinct, and the offspring shares exactly one allele with each
parent. As another example, any pair of individuals has one of three
possible ICCs, depending on whether they share 0, 1, or 2 alleles IBD,
with 2, 3, or 4 IBD-distinct alleles, respectively. The ICC
probabilities for an arbitrary noninbred set of relatives with a~given
relationship R can be computed using one of several software packages
[e.g., GENEHUNTER, described by Kruglyak et al. (\citeyear{Kruetal96})]. For a~set of
loci in a chromosomal region for which there is negligible probability
of meiotic recombination within a family, the diploid haplotype
probabilities for a family with a given ICC $\phi$ are obtained by
independently assigning a~haplotype to each of its IBD-distinct alleles
[see Whittemore and Halpern (\citeyear{WhiHal94})]. Table~\ref{tab5} shows probabilities of
the genotypes at a single locus for any pair of relatives, based on
their three possible ICCs.

{\renewcommand{\thetable}{\Alph{section}.\arabic{table}}
\setcounter{table}{0}
\begin{table}
\caption{Probability of genotypes $( g_{1},g_{2} )$
for a pair of individuals at a SNP with minor allele frequency $p
= 1 - q$, conditional on the number $\phi$ of alleles they share
IBD\protect\tabnoteref[a]{t51}}
\label{tab5}
\begin{tabular*}{\textwidth}{@{\extracolsep{\fill}}lccc@{}}
\hline
\multicolumn{1}{@{}l}{$\bolds{( g_{1},g_{2} )}$} & \multicolumn{1}{c}{$\bolds{\operatorname{Pr}(g_{1},g_{2}|\phi = 0 )}$}
&\multicolumn{1}{c}{$\bolds{\operatorname{Pr}( g_{1},g_{2}|\phi = 1 )}$} & \multicolumn{1}{c@{}}{$\bolds{\operatorname{Pr}( g_{1},g_{2}|\phi = 2 )}$}\\
\hline
$(0,0)$ & $q^{4}$ & $q^{3}$ & $q^{2}$\\
$(0,1)$ or $(1,0)$ & $4pq^{3}$ & $2pq^{2}$ & $0$\\
$(0,2)$ or $(2,0)$ & $2p^{2}q^{2}$ & $0$ & $0$\\
$(1,1)$ & $4p^{2}q^{2}$ & $p^{2}q+pq^{2}$ & $2pq$\\
$(1,2)$ or $(2,1)$ & $4p^{3}q$ & $2p^{2}q$ & $0$\\
$(2,2)$ & $p^{4}$ & $p^{3}$ & $p^{2}$\\
Total & $1$ & $1$ & $1$\\
\hline
\end{tabular*}
\tabnotetext[a]{t51}{Identity-by-descent.}
\end{table}}

B. \textit{Using the EM algorithm to maximize the likelihood function}
$L( \bolds{\theta})$ \textit{of equation \textup{(\ref{eq1})}.} As described by Dempster, Laird,
and Rubin (\citeyear{DemLaiRub77}), the algorithm involves iterative applications of an
expectation (E) step and a maximization (M) step, starting with an
initial parameter value $\theta^{_{0}} = ( \pi^{_{0}},\alpha^{_{0}} )$.
(Here we used HW genotype frequencies obtained\vspace*{1pt} from a MAF of 0.2 for
$\pi^{0}$ and a read error rate of 0.01 for $\alpha^{0}$.) At iteration
$t$, the E step computes the expectations of the family genotypes (or
diploid haplotypes), given the read data and the current parameter value
$\theta^{t - 1}$. Then $\theta^{t - 1}$ is updated to $\theta^{t}$ by
maximizing the complete data likelihood, given by
%
\renewcommand{\theequation}{A.\arabic{equation}}
\setcounter{equation}{0}
\begin{equation}
\label{eqA.1}\quad
\prod_{i} \Pr( G_{i},Y_{i}|N_{i},R;\theta) = \prod_{i}
\Pr( G_{i}|R;\pi) \prod_{i} \Pr(
Y_{i}|G_{i},N_{i};\alpha) ,
\end{equation}
with unobserved genotypes $G_{i}$ replaced by their conditional
expectations, calculated using the IBD configurations described in
the \hyperref[app]{Appendix}, Part~A. These two steps are repeated until the relative difference
between estimates~$\theta_{t - 1}$ and $\theta_{t}$ differ by no more
than a prespecified small amount (here we used $\frac{|\theta _{t - 1}
- \theta _{t}|}{\theta _{t - 1}} \le 10^{ - 8})$. If the procedure
produced a parameter on the boundary of the parameter space leading to
a local maximum, we randomly selected another initial parameter and
reran the algorithm.

We illustrate the procedure as applied to a set of $I$ noninbred
parent--offspring trios. We assume random parental mating and a single
read nucleotide error probability $\alpha$. In this case we can
parameterize as $\theta = ( \pi_{0},\pi_{2},\alpha)$, where $\pi_{g}$ is
the probability that a parent carries g copies of the variant allele, $g
= 0,1,2$, and $\pi_{1} = 1 - \pi_{0} - \pi_{2}$. The first factor on the
right side of (\ref{eqA.1}) is a data-dependent constant times the term
$\prod_{g = 0}^{2} \pi_{g}^{x_{g}}$, where $x_{g}$ is the total number
of parents with genotype $g$, with $x_{0}+x_{1}+x_{2} = 2I$. The second
factor is proportional to $\alpha^{u}( 1 - \alpha)^{n_{ +} - u}$, where
$n_{+}$ is the total number of reads among homozygous individuals (those
with $g=0$ or $g=2$), and $u$ is the number of incorrect reads among
these individuals (i.e., variant reads for those with $g=0$ or
nonvariant reads for those with $g=2$). Thus, the E-step\vspace*{1pt} involves
computing the expectations of $x_{0}$, $x_{2}$, $n_{+}$, and $u$,
conditional on~$\theta^{t - 1}$, the total number of reads, and the
total number of variant reads. The M-step involves setting
$\pi_{_{g}}^{t} = E[ x_{g} ]/2I, g=0,2$, and $\alpha^{t} = u/n_{ +}$.
\end{appendix}

\section*{Acknowledgments}

We thank Chiara
Sabatti, Douglas F. Levinson, and Wing Wong for helpful discussions.


\printaddresses

\end{document}